# Pinning and movement of individual nanoscale magnetic skyrmions via defects

Christian Hanneken, André Kubetzka, Kirsten von Bergmann, and Roland Wiesendanger

Department of Physics, University of Hamburg, 20355 Hamburg, Germany

An understanding of the pinning of magnetic skyrmions to defects is crucial for the development of future spintronic applications. While pinning is desirable for a precise positioning of magnetic skyrmions it is detrimental when they are to be moved through a material. We use scanning tunneling microscopy to study the interaction between atomic scale defects and magnetic skyrmions that are only a few nanometers in diameter. The studied pinning centers range from single atom inlayer defects and adatoms to clusters adsorbed on the surface of our model system. We find very different pinning strengths and identify preferred positions of the skyrmion. The interaction between a cluster and a skyrmion can be sufficiently strong for the skyrmion to follow when the cluster is moved across the surface by lateral manipulation with the STM tip.

Magnetic skyrmions are particle-like knots in the spin texture which are considered as promising candidates for future spintronic applications [1-3]. They typically arise in materials which have a spin spiral ground state due to the Dzyaloshinskii-Moriya (DM) interaction [4,5] at zero magnetic field; upon application of an external magnetic field a hexagonal skyrmion lattice phase can occur before the magnetization is saturated. From an analysis of the energies it is known that near the boundary between the lattice phase and the ferromagnetic state also isolated skyrmions can be stabilized, and it has been demonstrated experimentally that also reduced temperature facilitates the preparation of individual skyrmions [6]. Skyrmion lattices have been moved through a material by laterally imprinted currents due to spin transfer torques [7-9]. While below a critical current density the periodic magnetic texture is pinned, for larger lateral currents pinning forces were found to be negligible and the skyrmion velocity was proportional to the current [8]. Simulations for skyrmion lattices support this finding and suggest that randomly distributed impurities do not have a significant impact on the velocity under applied currents [10]; however, a dependence on the density of magnetic skyrmions on their lateral movement has been proposed [11]. For individual magnetic skyrmions several different pinning mechanisms have been considered theoretically, ranging from combined increase of magnetic exchange and DM interaction [12], over vacancies in the magnetic material [13], to repulsive interaction due to areas with higher magnetic anisotropy [14]. The conclusion from these studies is that depending on the parameters both the movement around a defect and the capturing of a skyrmion at a pinning site is possible.

Here we study pinning in the model system of the PdFe atomic bilayer on Ir(111) [6,15,16]. Using scanning tunneling microscopy (STM) we investigate the interaction of non-collinear magnetic states with different atomic-scale defects. We find that single magnetic skyrmions can be significantly distorted due to their interaction with clusters adsorbed on the surface. Furthermore the lateral movement of skyrmions via local manipulation of the pinning cluster by the tip of the STM is possible.

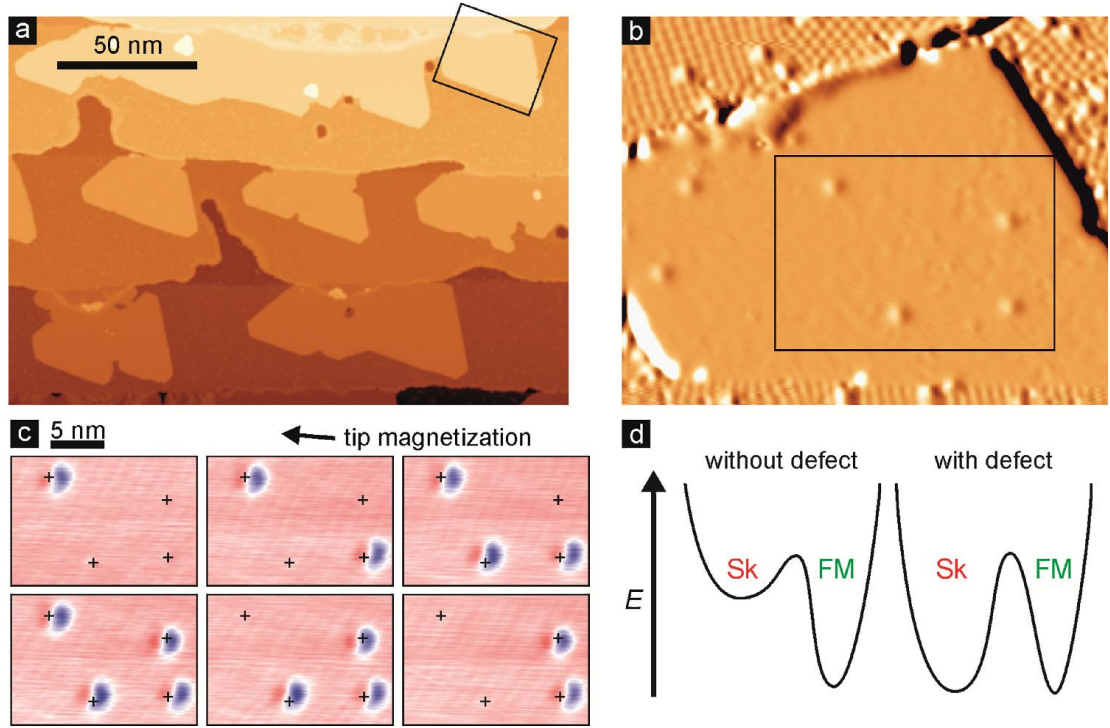

Figure 1: Strongly pinning inlayer defects. (a) Overview constant-current STM image of about 0.3 atomic layers of Pd on about 0.8 atomic layers of Fe on Ir(111). (b) Derivative of a constant-current STM image of the area indicated by the box in (a); in the Pd layer four single atom inlayer defects are observed within the box while in the Fe layer the magnetic nanoskyrmion lattice is present. (c) Magnetic contribution to SP-STM constant-current images of the area indicated by the box in (b) for a different number and arrangement of pinned magnetic skyrmions; the positions of the defects are indicated by crosses. The images are obtained from constant-current images measured with a tip magnetization as indicated by subtracting a calculated spin-averaged reference STM image; same dataset as in [6]. (d) Sketches of double well potential of skyrmion (Sk) and ferromagnetic (FM) states; at identical magnetic fields the presence of a defect leads to a decrease of the skyrmion energy with respect to the ferromagnetic state. (Measurement parameters: $B$ = +3.25 T, $T$ = 4.2 K, Cr tip sensitive to the in-plane sample magnetization, (a): $U$ = +0.1 V, $I$ = 0.5 nA, (b,c): $U$ = +0.25 V, $I$ = 1 nA).

The PdFe atomic bilayer on Ir(111), see overview STM image in Fig. 1(a), has been studied before by spin-polarized (SP-) STM [17,18] and it was found that this system exhibits an interface-induced DM spin spiral at zero magnetic field; when an external magnetic field is applied a transition to a skyrmion lattice phase and finally to the ferromagnetic state is induced [6]. At a magnetic field of $B$ = +3.25 T, as in Fig. 1, the system is in the ferromagnetic phase for the pristine PdFe layer. However, it has been observed that skyrmions can survive and also be generated in the vicinity of inlayer defects [6], such as the four impurity atoms in the Pd layer shown in the box of the differentiated SP-STM image of Fig. 1(b). We assume that they are single Fe atoms within the Pd layer as a result of intermixing. Figure 1(c) shows the in-plane components of the sample magnetization along the tip magnetization axis for several subsequent SP-STM measurements. Due to the in-plane tip magnetization the axially symmetric skyrmions are imaged as two-lobe structures [16] and in between the images they were written or deleted with the STM tip [6]. The fact that all skyrmions appear in the same way, even after deleting and rewriting, shows that they have the same rotational sense due to the DM interaction. At this magnetic field they are typically found in direct vicinity to the defects, indicated by the crosses. From this we conclude that these single inlayer defects modify the potential landscape and lower the energy of the skyrmion state with respect to the ferromagnetic state, see sketch in Fig. 1(d). Such a modification of the energy landscape has been suggested before based on simulations with defects of increased exchange interaction [12]. Interestingly the defects are located slightly off-center with respect to the axially-symmetric skyrmions, see Fig. 1(c). In principle this should lead to a situation where the skyrmion center is free to move around the defect. While we observe different orientations of the skyrmion center for different defects, for a single defect the skyrmion is always locked to the same position, compare the different images in Fig. 1c. This indicates that the exact position is not only determined by the single inlayer defects, but also by subsurface defects, generating a spatially inhomogeneous environment.

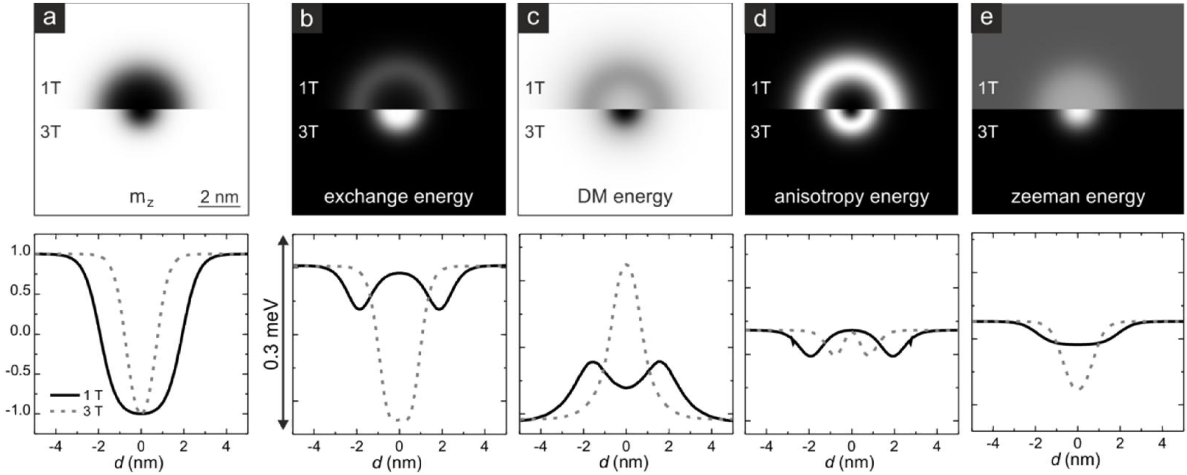

Figure 2: Pinning as a function of material parameters. (a) Out-of-plane magnetization components of a skyrmion in PdFe at $B$ = 1 T and 3 T, upper and lower half of image, respectively; the plot at the bottom shows cuts through the skyrmion center for the two magnetic field values. (b) Magnetic exchange stiffness density, the plot at the bottom shows the energy change due to the exchange stiffness as function of defect position across the skyrmion. (c) DM interaction, (d) magnetocrystalline anisotropy, and (e) Zeeman energy density, respectively, displayed in analogous fashion to (b). The defect is modelled as a Gaussian with 10% reduced value and full width at half maximum of 0.235 nm.

Different origins for pinning at a defect are possible, and an analysis of the different energies relevant for skyrmion formation gives a qualitative picture. For hcp-stacked Pd on fcc-stacked Fe the material parameters have been determined experimentally: exchange stiffness $A$ = 2.0 pJ/m, DM interaction $D$ = 3.9 mJ/m$^2$, and anisotropy $K$ = 2.5 MJ/m$^3$ [16]. Figure 2(a) shows the spatially resolved out-of-plane magnetization components of a magnetic skyrmion at $B$ = 1 T and $B$ = 3 T (upper and lower halves of image, respectively) as well as cuts through the skyrmion center as obtained from the energy functional [16]. The respective exchange stiffness densities are shown in Fig. 2(b) in an analogous fashion: the ferromagnetic environment outside the skyrmion maximizes the energy gain due to the ferromagnetic exchange interaction, and the bright ring/dot indicates the region in the skyrmion which is most unfavorable for this interaction. For the comparably large skyrmion at 1 T the spin configuration in the skyrmion center is again more favorable for ferromagnetic exchange. When we introduce an atomic-scale defect with smaller exchange stiffness and calculate the energy of the system as a function of the lateral position of the defect across the skyrmion, we arrive at the plot shown at the bottom of Fig. 2(b): for such a defect the energy is minimized when it is located at the maximum of the exchange energy density, i.e. the bright region in the upper image of Fig. 2(b). For the 1 T skyrmion also a defect in the center of the skyrmion is unfavorable. This means that an inlayer defect with reduced exchange interaction would favor a situation where the skyrmion is pinned off-center with respect to the defect position at low magnetic fields but it is pinned in the center for higher field values. For a defect with increased exchange stiffness the laterally-resolved energy plot due to the interaction with the defect (Fig. 2(b) bottom) is inverted.

The energy gain due to DMI is maximized for a skyrmion at $B = 1$ T in a ring around the center, where a canting between adjacent moment occurs, see Fig. 2(c) top. For 3 T the DMI energy density shows a dot. Accordingly, a defect with decreased DMI would be found either outside the skyrmion or, for $B = 1$ T, it could also be in its center, cf. Fig. 2(c) bottom. The anisotropy density is directly correlated to the in-plane component of magnetization, Fig. 2(d) top, giving rise to a defect position in the in-plane region for decreased out-of-plane anisotropy, Fig. 2(d) bottom. A decrease of the magnetic moment of a defect compared to the PdFe affects the Zeeman energy density, cf. Fig. 2(e). This decomposition into the different energy contribution for a magnetic skyrmion enables a general understanding on the pinning properties for the decrease or increase of an interaction at the defect position. We find that an off-center defect configuration as observed in Fig. 1(c) can be due to either decreased $A$ or increased $D$ for large skyrmions, or a smaller $K$ of the defect with respect to the environment for any skyrmion size.

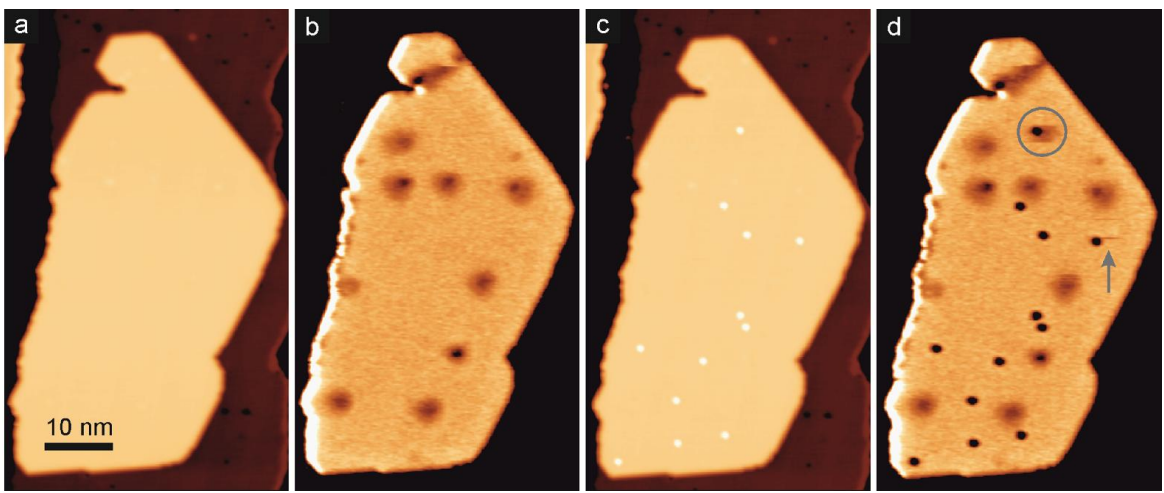

Figure 3: Single Co atoms adsorbed on the PdFe bilayer. (a) Constant-current image of one Pd island on the Fe monolayer and (b) simultaneously acquired d$I$/d$U$ map showing several skyrmions that are pinned to inlayer defects; one spin spiral fragment is observed at a constriction at the top; the imaging mechanism is NCMR. (c) Constant-current image of the same PdFe island after deposition of single Co atoms and (d) corresponding d$I$/d$U$ map; one additional skyrmion has appeared, see circle, and one skyrmion is frequently written and deleted, see arrow. (Measurement parameters: $B = +1.8$ T, $T = 8$ K, Cr tip, $U = +0.7$ V, $I = 1$ nA).

While the native inlayer defects observed in Fig. 1 show strong pinning one cannot manipulate them with the STM tip and a bottom-up fabrication of skyrmion arrays is not possible. To be able to manipulate atoms they need to be adsorbed on top of the surface. To investigate the pinning properties of such adatoms we compare the same sample area before and after deposition of single Co atoms. Figure 3(a) shows the topography of a Pd island on the Fe monolayer and (b) the corresponding d$I$/d$U$ map taken at +0.7 V. This sample bias voltage was chosen for two reasons: first, the energy of the injected electrons is high enough to be able to switch between skyrmionic and ferromagnetic state [6] and thus we assume that the favorable magnetic state is observed. Second, this sample bias voltage enables the detection of non-collinear magnetic states in PdFe even with a non-magnetic tip due to non-collinear magnetoresistance (NCMR) [19]: due to spin mixing the density of states depends on the angle between nearest-neighbor magnetic moments, and thus a skyrmion is electronically different from the ferromagnetic state, giving rise to a signal change in d$I$/d$U$ maps taken at +0.7 V.

In Fig. 3(b) the skyrmions are imaged as dark dots and we again observe that they are pinned to inlayer defects. Since the sample preparation was comparable and the defects are again off-center with respect to the skyrmions, we assume that the nature of the inlayer atoms is the same as in Fig. 1, i.e. Fe atoms embedded in the Pd layer. To investigate the influence of single Co adatoms they were deposited *in situ* into the cold STM ($T < 20$ K). The resulting topographic image of the same PdFe island now shows several protrusions, Fig. 3(c), which indicate the positions of single Co atoms. In the corresponding d$I$/d$U$ map of Fig. 3(d) the single Co atoms are imaged as dark spots, and it becomes obvious that they do not have a strong influence on the magnetic skyrmions that are pinned to the inlayer defects. Only one additional magnetic skyrmion is observed at a position of a Co adatom, see circle, and one skyrmion is frequently generated and annihilated at a different one, see arrow. This shows that single Co adatoms do not have a significant impact on the potential landscape for skyrmionic states in PdFe, possibly because most of the magnetic moment is located at the Fe atoms which are separated from the Co adatoms by the Pd layer, inhibiting an efficient change of local material parameters.

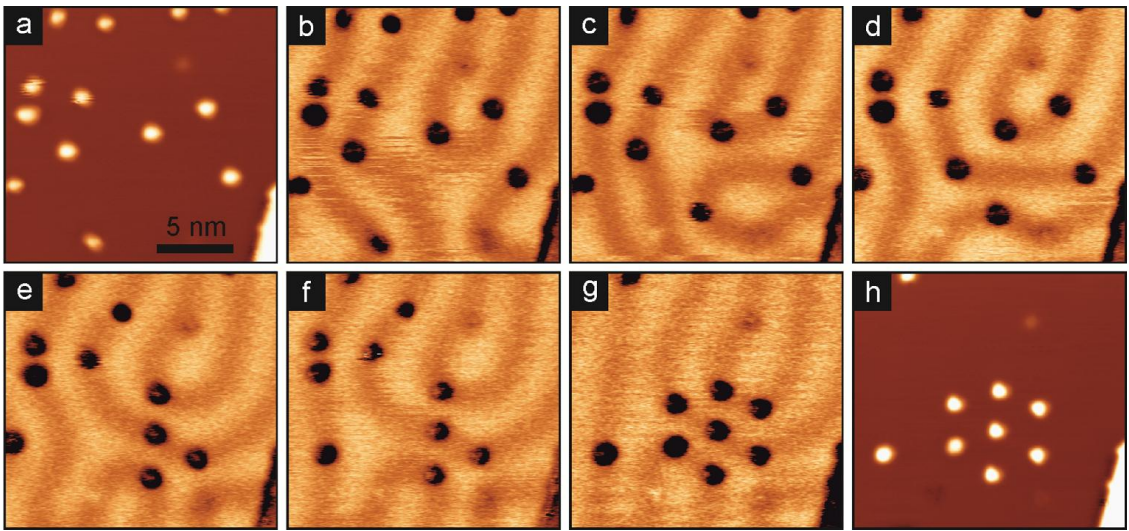

Figure 4: Pinning of the spin spiral to Co clusters. (a)-(h) Series of images showing the response of the spin spiral state to the position of Co clusters of various sizes. (b)-(g) are d$I$/d$U$ maps with NCMR contrast, and (a) and (h) are constant-current images corresponding to (b) and (g), respectively. (Measurement parameters: $B = 0$ T, $T = 4.2$ K, Cr tip that changed during this series, $U = +0.7$ V, $I = 0.3$ nA; manipulation parameters: $U = +1$mV to $+6$mV, $I = 35$ nA to 75 nA).

When Co is deposited in situ at a slightly higher temperature we find Co clusters on the PdFe surface, see Fig. 4. Comparison with bottom-up fabricated clusters indicates that typically Co-dimers and Co-trimers are formed by self-organization. The measurements of Fig. 4 are performed in the absence of an external magnetic field and the imaging mechanism of the spin spiral state is again due to NCMR [19]: since the PdFe system has an out-of-plane anisotropy [15,16] the spin rotation, i.e. the degree of non-collinearity, is larger in in-plane magnetized areas as compared to out-of-plane easy axis spins. For our system this results in a smaller d$I$/d$U$ signal for sample areas with in-plane components of the spin spiral as compared to out-of-plane regions, leading to a period in the NCMR that is half of the magnetic period.

The protrusions on the PdFe layer in Fig. 4 are mostly Co trimers, and one can observe that there is a correlation between the position of the clusters and the darker NCMR contrast, i.e. the in-plane part of the spin spiral. When the clusters are moved by lateral manipulation with the STM tip, the spin spiral rearranges to again have its in-plane part at the position of the clusters. This shows that while single Co adatoms were found to have a negligible influence on the presence of skyrmions, Fig. 3, we observe that clusters of Co atoms effectively interact with the spin spiral state.

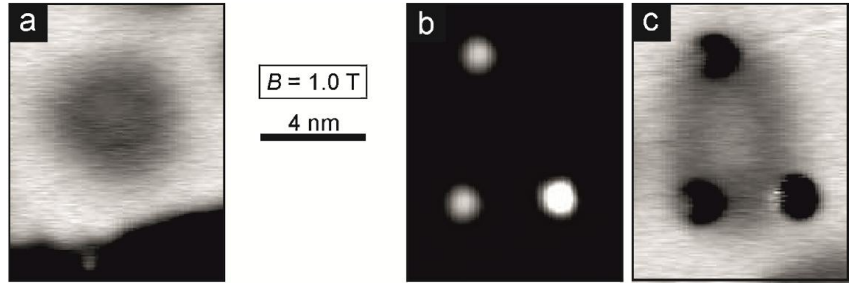

Figure 5: Distortion of single skyrmions. (a) d$I$/d$U$ map of an unperturbed magnetic skyrmion at $B$ = 1 T. (b),(c) Constant-current image and corresponding d$I$/d$U$ map of a skyrmion that is pinned to three clusters of different sizes. (Measurement parameters: $B$ = +1.0 T, $T$ = 4.2 K, Cr tip showing NCMR contrast, $U$ = +0.7 V, $I$ = 1 nA).

This strong pinning to clusters is also observed for magnetic skyrmions, see Fig. 5. While unperturbed isolated skyrmions at $B$ = +1 T are typically imaged as shown in Fig. 5(a), in the presence of clusters they can be strongly distorted and enlarged as in the example of Fig. 5(b),(c): here the aspect ratio of vertical and horizontal dimension is about 1.75.

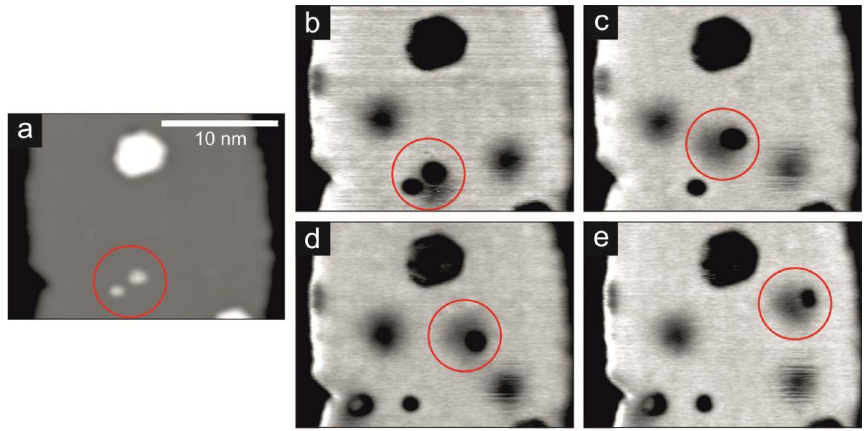

Figure 6: Lateral movement of a single magnetic skyrmion. (a) Constant-current image showing two clusters (red circle) on a PdFe stripe, and a Pd island (upper part). (b)-(e) Sequence of d$I$/d$U$ maps with three skyrmions (NCMR contrast); in between the images one cluster (red circle) has been manipulated over the surface with the STM tip. (Measurement parameters: $B$ = +2.1 T, $T$ = 8 K, Cr tip, $U$ = +0.7 V, $I$ = 1 nA. Manipulation parameters: $U$ = +1 mV to +6 mV, $I$ = 35 nA to 75 nA. The tip apex changed several times during this image series.).

Figure 6(a) shows the topography of a PdFe stripe with a small second layer Pd island at the top. The two small protrusions at the bottom have a different size and while the left one is most likely a single Co adatom, the right one is probably a trimer which contains also some Cr from the tip. The d$I$/d$U$ map of Fig. 6(b) shows that there is a skyrmion at the cluster and two other skyrmions nearby, which are pinned to inlayer defects. The cluster was moved repeatedly by the STM tip, Fig. 6(b)-(e). For the lateral movement of the cluster a low sample bias of only +1 mV to +6 mV was used. Since the switching probability of the magnetic state becomes negligible in this small voltage regime [6], we expect that we do not delete the magnetic skyrmion, despite of the large currents required for atom manipulation. Indeed, after each manipulation step the skyrmion is found at the cluster for each new position, cf. Fig. 6(b)-(e). This demonstrates that the pinning of the skyrmion to the cluster is strong enough to keep it in close vicinity while moving the cluster across the surface. This pinning can again be understood within the double well potential model sketched in Fig. 1(d) and the energy density plots of Fig. 2, where now the skyrmion minimum in the vicinity of the defect is moved across the surface, keeping the magnetic state intact.

In conclusion, we have shown that defects in and on top of the PdFe bilayer locally modify the potential landscape. Inlayer defects, which are presumably single Fe atoms in the Pd layer, interact strongly with single magnetic skyrmions and are typically found off-center with respect to the radially-symmetric spin configuration. Their presence enables the creation of skyrmions at magnetic fields values that favor the ferromagnetic state, demonstrating a change of the alignment of the relative energy levels. The disentanglement of the spatial distribution of competing energies for skyrmion formation facilitates a general understanding of possible causes for pinning. While single Co adatoms on top of the magnetically polarized Pd layer do not influence the magnetic texture significantly, clusters of Co atoms can be used to modify the details of a spin spiral state. In addition, skyrmions can be distorted due to pinning at more than one Co cluster. The pinning to clusters is also sufficiently strong to move an individual skyrmion laterally, as demonstrated by local manipulation of the pinning cluster with the STM tip. These findings demonstrate that a tuning of pinning properties is feasible, making it possible to design structures of pinning sites that define preferred positions for magnetic skyrmions.

## Acknowledgements

We thank Niklas Romming, Bertrand Dupé, and Stefan Heinze for discussions and acknowledge financial support from the German Research Foundation (DFG: GrK 1286 and SFB 668-A8) and the European Union (FET-Open project MAGicSky No. 665095).